\documentclass[aps,prb,twocolumn,groupedaddress]{revtex4}
\usepackage{longtable}
\usepackage{graphicx}% Include figure files
\usepackage{dcolumn}% Align table columns on decimal point
\usepackage{bm}% bold math
\newcommand{\PCCO}{Pr$_{2-x}$Ce$_x$CuO$_{4-\delta}$}
\newcommand{\QPT}{quantum phase transition}
\newcommand{\QCP}{quantum critical point}
\newcommand{\QC}{quantum critical}
\newcommand{\FLQ}{Fermi liquid}
\newcommand{\FS}{Fermi surface}
\newcommand{\NCCO}{Nd$_{2-x}$Ce$_x$CuO$_4$}
\newcommand{\AFM}{antiferromagnetic}
\newcommand{\MR}{magnetoresistance}
\newcommand{\nMR}{negative magnetoresistance}

\newcommand{\etal}{\emph{et al.}}
\newcommand{\neel}{N\`{e}el}

\begin{document}

\title{Fermi Surface Reconstruction in the Electron-doped Cuprate Pr$_{2-x}$Ce$_x$CuO$_{4-\delta}$;\\
A transport study}


\author{Y. Dagan}
\email[]{yodagan@post.tau.ac.il} \affiliation{Raymond and Beverly Sackler School of Physics
and Astronomy, Tel-Aviv University, Tel Aviv, 69978, Israel}
\author{R.L. Greene}
\affiliation{Center for Nanophysics and Advanced Materials, Physics
Department, University of Maryland College Park, MD, 20743}


\date{\today}

\begin{abstract}
We report extensive resistivity, Hall, and magnetoresistance measurements on thin films of the electron-doped cuprate \PCCO~(PCCO), as a function of doping, temperature and magnetic field. The doping dependence of the resistivity and Hall number at low temperatures are characteristic of a system near a quantum phase transition or a Fermi Surface Reconstruction (FSR) point.  The spin magnetoresistance drops to zero near the critical point. The data presented in this paper were compiled during the 2004-2007 period but were never published in this comprehensive form. Because of the recent interest in very similar results now being found in the normal state of hole-doped cuprates, we believe the results of our older, mostly unpublished, work will be of interest to the present community of cuprate researchers. In particular, Fig.\ref{nH} shows the large change in Hall number at the FSR point in \PCCO, similar to that found recently in YBa$_2$Cu$_3$O$_{7-\delta}$ and La$_{2-x}$Sr$_x$CuO$_4$ (See ~\cite{badoux2016change} and ~\cite{collignon2016fermi}). Also, Fig.\ref{Upturn} illustrates how the resistivity upturn is affected by the FSR. The cause of the resistivity upturn has been attributed to the loss of carriers at doping below the FSR in the hole-doped cuprates (see \cite{laliberte2016origin}), however, this scenario does not explain the data for \PCCO. The upturn in n-doped cuprates is more-likely due to a combination of carrier decrease and a change in the scattering rate below the FSR \cite{finkelman2010resistivity}. The change in spin scattering below the FSR is illustrated by Fig.\ref{dopdepofSMR} in this paper. Chen \etal~ \cite{chen2009theory} have developed a model based on spin scattering that is able to explain qualitatively the resistivity upturn in all the cuprates.
\end{abstract}

%\pacs{74.25.Fy, 71.10.Hf, 73.43.Nq, 74.72.-h, 73.43.Qt }
%\keywords{}

\maketitle
\section {Introduction}
The electron doped (n-doped) cuprates,
(RE$_{2-x}$Ce$_x$CuO$_{4-\delta}$ with RE=Nd, Pr, La, Sm),
superconductors offer a unique system for studying the low
temperatures normal state properties of a high T$_c$ cuprate. In
most of the high T$_c$ cuprates a very high field, usually
inaccessible, is needed to quench superconductivity. The normal
state is thus obscured by the occurrence of superconductivity. In
the n-doped cuprates the normal state is accessible at modest
magnetic fields $(H<10T)$. Moreover, there is now compelling
evidence that a quantum phase transition occurs as a function of
doping slightly above optimum doping: the normal state Hall
coefficient of \PCCO~ at 350mK exhibits an abrupt change at
$x=0.165\pm0.005$ \cite{daganResistivityPRL}. This possibly
singular behavior is accompanied by significant changes in the
temperature dependence of the normal state resistivity below 20K.
It was also found that the spin related magnetoresistance suddenly
vanishes for $x\geq0.16$ at T=1.5K. This \MR~ appears in the
region of the doping-temperature phase diagram where $d\rho/dT<0$,
or an upturn in the resistivity appears, thus suggesting that the
upturn partially originates from spin scattering processes that
vanish at the \QCP~ \cite{daganMR}. The broad antiferromagnetic
region from x=0 to just above x=0.15 found in the phase diagram of
the n-doped cuprates
\cite{LukeAFM}$^,$\cite{KangAFM}$^,$\cite{fujita:147003}$^,$\cite{Skanthakumar}
suggests that the \QCP~ found in ref.\cite{daganResistivityPRL}
can be associated with the disappearance of the \AFM~ phase as the
doping is increased at T=0.
\par
Tunneling measurements give evidence for a normal state tunneling
gap for dopings within the superconducting dome
\cite{KleefischNSG}$^,$\cite{biswasNSTG} this gap was reported to
vanish near x=0.17.\cite{alff2003hidden} However, Dagan
\etal~\cite{daganQazilbashtunneling} reported that this gap
persists up to x=0.19 and for overdoped samples its behavior
follows the doping dependence of T$_c$.
\par
Extensive optical measurements of \PCCO~ as a function of doping
revealed the signature of a density wave gap that vanishes above a
critical doping,
$x_c>0.15$.\cite{Onose}$^,$\cite{zimmersEPL}$^,$\cite{zimmers2007infrared}
Consistent with the \QCP~ found from transport
 \cite{daganResistivityPRL} and with the neutron scattering
measurements.\cite{LukeAFM}$^,$\cite{KangAFM}$^,$\cite{fujita:147003}$^,$\cite{Skanthakumar}
Pengcheng Dai \etal~\cite{daiAFMoxygenvariation} studied the
electron-doped compound Pr$_{0.8}$LaCe$_{0.12}$CuO$_{4-\delta}$ as
a function of the oxygen content $\delta$ by means of neutron
scattering. They found that the \neel~ temperatures of the three
dimensional ordered antiferromagnetism and of a quasi two
dimensional spin density wave modulation vanish close to optimum
oxygen content. Dai \etal~\cite{daiAFMoxygenvariation} suggested
that superconductivity and antiferromagnetism coexist near a
\QCP~. By contrast Motoyama \etal~\cite{GrevenQCP013} have found
that the spin correlation length remains finite at doping levels
$x\geq0.145$ while at low temperatures it diverges near
$x\approx0.13$ where superconductivity onsets. From these results
they concluded that superconductivity and antiferromagnetism do
not coexist but are separated by a quantum phase transition at
$x\approx0.13$. Angular magnetoresistance measurements \cite{Yu_nMR} were found to be consistent with ref.\cite{daganResistivityPRL}. Recent quantum oscillations measurements on Nd$_{2-x}$Ce$_{x}$CuO$_{4-\delta}$ found an abrupt change in the oscillation frequency \cite{GrossQOscinNCCO} at the same doping level found in Ref.\cite{daganResistivityPRL}. The discrepancy between the transport measurements and the recent neutron scattering data regarding the location of the quantum critical point may be reconciled using a recent contribution suggesting that due to the occurrence of superconductivity the quantum critical point is shifted towards lower doping levels. When superconductivity is muted by a magnetic field the QCP shifts back to higher dopings.\cite{sachdevRMP}
\par
Angle resolved photoemission spectroscopy (ARPES) experiments on
\NCCO~ show a remarkable change in the \FS~ as a function of
doping, going from a small electron pocket at $(\pi,0)$ in
underdoped (x=0.04) to a large hole-like \FS~ at optimal doping
(x=0.15) \cite{armitageprldoping}$^,$\cite{matsui:047005}. The
normal state transport properties of the cuprates for T$>$T$_c$ do
not follow the behavior expected for conventional metals (\FLQ~).
For example, the Hall coefficient has a strong temperature
dependence and the ab-plane resistivity varies as T (for p-type),
T$^2$ (for n-type) up to temperatures greater than 250K.
\par
Evidence for a \QPT~ in the hole-doped (p-type) cuprates has been
reported\cite{Loramtallon}$^,$\cite{BalakirevHall}$^,$\cite{panagopulus}.
However, there are major differences between the hole-doped and
the n-doped cuprates: a) the phase diagram of the electron-doped
(n-type) cuprates shows an antiferromagnetic phase that starts at
x=0 and persists up to, possibly into, the superconducting dome,
b) the transport properties are significantly different between
these two types of cuprates as will be elaborated below, c) the
pseudogap phase \cite{Timuskandstat} of unknown origin that is
found in underdoped hole-doped samples is apparently absent on the
electron-doped side. Theoretical work explained the differences in
magnetic behavior between p-doped and n-doped cuprates by weaker
electrons correlations, which are doping dependent in the latter
\cite{kyung:147004}.
\par
In various \QC~ models there are definite predictions for the
behavior of the resistivity and the Hall effect at temperatures
above the \QCP~ and for doping concentrations near x$_c$
\cite{ChakravartyHall}$^,$\cite{Rosch}$^,$\cite{RoschPRB}$^,$\cite{linandmillis}$^,$\cite{fentonmetalicQCP}.
Evidence for some of these predictions have been found in other
correlated systems such as heavy fermions
\cite{MathurQCP}$^,$\cite{SidorovQCP}. In the cuprates, other
suggestive evidence for a \QPT~ at an $x_c$ under the
superconducting dome has come from the observation of a low T
normal state "insulator" to metal crossover as a function of
doping in both n-and p-type materials
\cite{BoebingerANDOMIT}$^,$\cite{fournierPRL}.
\par
In this paper we review the results from references
\cite{daganResistivityPRL}$^,$\cite{daganMR} and present new data:
Hall and resistivity measurements in extended doping and
temperature ranges, and field and temperature dependences of the
orbital and spin \MR~. We discuss the new data in the framework of
the \QCP~ scenario and the other available models.

\section{Samples preparation and characterization.}
\begin{figure}
\includegraphics[width=1\hsize]{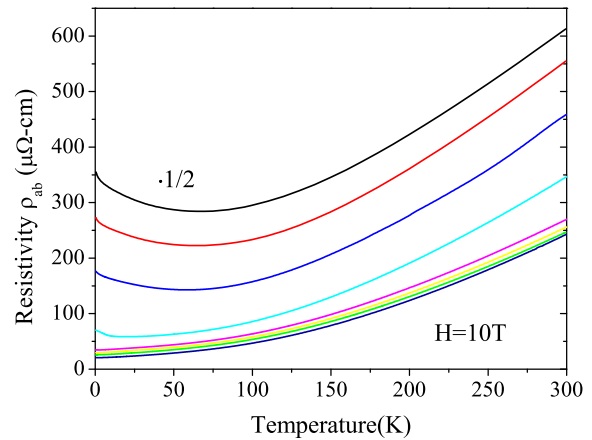}
\caption{(color online) The \textit{ab} plane resistivity as a
function of temperature for the various \PCCO~ samples (top to
bottom curves) $x=0.11$ (black) (the resistivity of this curve is
divided by two), $x=0.12$ (red), $x=0.13$ (blue), $x=0.15$ (cyan),
$x=0.16$ (magenta), $x=0.17$ (yellow), $x=0.18$ (green), and
$x=0.19$ (navy). Superconductivity is muted by applying a magnetic
field of 10 Tesla, perpendicular to the \textit{ab}
plane.}\label{NSresistivity}
\end{figure}
\begin{figure}
\includegraphics[width=1\hsize]{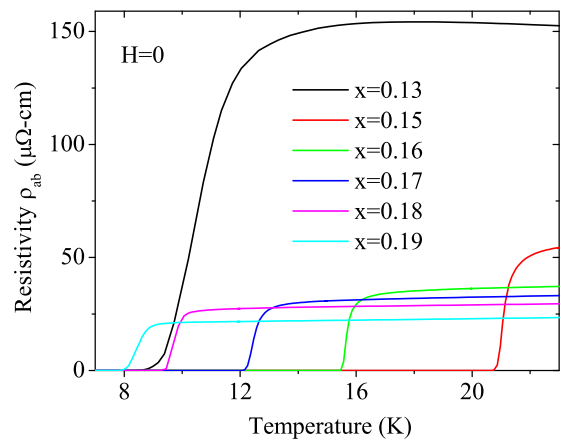}
\caption{(color online) The resistivity as a function of
temperature for the various doping levels at zero
field.}\label{SCresistivity}
\end{figure}
\par
\PCCO~ \textit{c}-axis oriented films of various cerium doping
concentrations were deposited from stoichiometric targets on (100)
oriented SrTiO$_3$ substrates using the pulsed laser deposition
technique with conditions similar to those reported by Maiser
\etal~\cite{Meisergrowthcond}. The films' thickness is
approximately 3000\AA. Rutherford back scattering measurements
were used to determine the thickness of the films. The
minimum-channeling yield obtained was 7-20 percents indicating a
good epitaxial growth. The low residual resistivity (see Figure
\ref{NSresistivity}) and sharp T$_c$ (see Figure
\ref{SCresistivity}) indicate that the films are of better quality
than the best previously reported pulse laser deposition films
\cite{fournierPRL} and comparable to molecular beam epitaxy grown
films \cite{Naitogrowthcond}. Since the oxygen content has an
influence on both the superconducting and normal state properties
of the material
\cite{WuJiang}$^,$\cite{higgins:104510}$^,$\cite{daiannealingnaturematerials}
we took extra care in optimizing the annealing process for each Ce
concentration. For the overdoped samples $(x > 0.15)$ annealing at
720$^\circ$C and pressure of 30 micro-Torr results initially in a
fast increase of T$_c$. Upon further annealing the films, both the
transition temperature and width decrease with time until
approaching a saturation point at time,$t_A$, where no changes in
these parameters are detected. Annealing for much longer time, $t
\gg t_A$, may result in material decomposition, holes of about 1
micron size are detected on the film's surface. We used $t_A$ as
the annealing time for our films. We found that $t_A$ increases
with increasing Ce doping and that roughly $t_A \propto d^2$, with
$d$ being the films' thickness, as expected for a diffusion
process. Since oxygen is diffusing, its travel distance is
proportional to the square root of time. Taking that distance as
the thickness of the film, $d$, one gets $d \propto \sqrt{t_A}$ or
$t_A \propto d^2$. For optimally doped and underdoped \PCCO~ we
had to simply maximize T$_c$ and avoid decomposition spots
detectable in an optical microscope. $t_A$ is typically 15 minutes
for optimally doped samples. For the non-superconducting sample
$x=0.11$ film we used the same annealing conditions as for
$x=0.13$. The saturation of the resistivity and critical
temperature as a function of annealing time at $t_A$ suggests that
the film reached equilibrium with the chambers atmosphere. By
maintaining constant annealing pressure and temperature we keep
similar oxygen content in all our films. Moreover, it has been
recently established that oxygen is a dopant (working in opposite
way to Ce) and causes disorder in the sample
\cite{higgins:104510}. The systematic dependence of T$_c$, $R_H$
and of $\rho(T)$ on the Ce doping suggests that the oxygen content
in all these films is indeed similar since all these quantities
are very sensitive to the oxygen content. The films were patterned
using a mechanical mask and ion milling technique to form Hall
bars 6.5mm long and 0.5mm wide.
\section{Measurements.}
Resistivity and Hall voltage were measured with a Quantum Design
"PPMS" system up to a magnetic field of 14T and down to a
temperature of 0.35K. The field was applied perpendicular to the
plane of the film, i.e. parallel to the \emph{c}-axis. At low
temperatures the Hall coefficient was measured by taking field
scans from -14T to 14T, a field at which the \PCCO~ is well in the
normal state even at 0.35K. Any \MR~ component from misalignment
of the Hall bar shape was removed by subtracting -H from +H data
in the usual way. At T$_c<$T$<$100K a narrower field range was
used. Above 100K the temperature was scanned at $\pm$14T and the
Hall voltage was calculated from the difference between these
measurements. When needed the field was aligned parallel to the ab
planes (H$\parallel ab$) with an accuracy better than
0.25$^\circ$. Measurements in the National High Magnetic Field Lab
(NHMFL) were taken in a 32.4 T resistive magnet and at
temperatures ranging from 1.5 K to 20 K. To exclude eddy current
heating effects we ensured that the data was reproducible,
symmetric for positive and negative magnetic fields and
independent of the sweeping rate. We measure the ab-plane
resistivity with a standard 4-probe technique.
\section{Results and discussion.}
\subsection{\label{resistivity} Resistivity}
\par
In Figure \ref{NSresistivity} we show the \emph{ab}-plane
resistivity \emph{versus} temperature at 10T $(H>H_{c2})$ for
$0.11\leq x \leq 0.19$. First, we note the decrease in the
resistivity as the Ce concentration is increased. Another feature
appearing in $x \leq 0.15$ films is a sign change in $d \rho/dT$
(an upturn). The temperature at which the upturn appears decreases
with increasing doping. Figure \ref{SCresistivity} shows the
resistive superconducting transition where T$_c$ has the expected
doping dependence \cite{Meisergrowthcond}. All the films have
sharp transitions. The transition width, $\Delta$T$_c$, measured
as the width at half maximum of the peak in $d \rho/dT$, is:
$\Delta$T$_c$=0.3-0.6 K in optimum and overdoped \PCCO~
(increasing with increasing Ce doping), and $\Delta$T$_c$=2.2 K
for x=0.13, all much sharper than previously reported pulse laser
deposition films.
\par

\begin{figure}
\includegraphics[width=1\hsize]{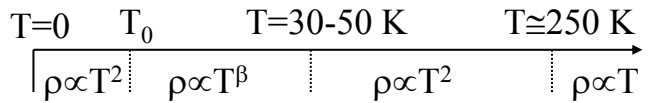}
\caption{Schematic illustration of the various regions in
temperature where different resistivity behavior is observed for
overdoped samples.}\label{resistivityregions}
\end{figure}

\begin{figure}
\includegraphics[width=1\hsize]{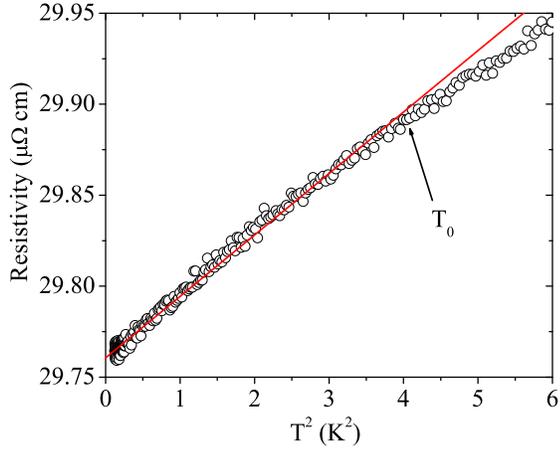}
\caption{The resistivity as a function of T$^2$ for the $x=0.17$
sample (circles) the line is a linear fit. Below $T_0\simeq2K$ a
T$^2$ behavior is observed.}\label{FLbehavioratLT}
\end{figure}

At the high temperatures (above 250K) the resistivity $\rho(T)$ is
roughly linear with temperature. In the intermediate temperature
range (from say 30-50K to around 200K) $\rho$ is proportional to
T$^2~$ \cite{Onose}. We note that this behavior is not the usual
Fermi liquid behavior expected at much lower temperatures. As the
temperature is decreased from around 30 K the temperature
dependence of the resistivity changes again from T$^2$ to
T$^\beta$  (with $2>\beta
>1$) and then to T$^2$ again at the lowest temperatures (below
T$_0(x)$). The different temperature regions are schematically
illustrated in Figure \ref{resistivityregions}. The low
temperature region is the focus of this section.
\par
\begin{figure}
\includegraphics[width=1\hsize]{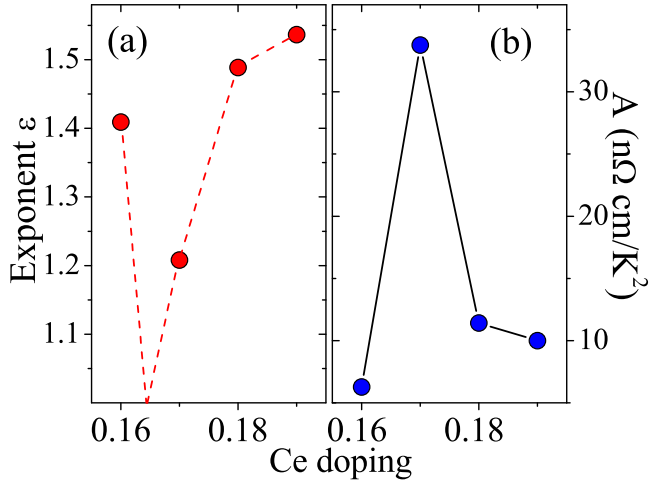}
\caption {a. The resistivity exponent $\varepsilon$ as found from
a fit to $\rho(T)=\rho_0+CT^{\varepsilon}$ for $0.35K>T>20K$ (the
line is a guide to the eye). b. The coefficient $A$ of the T$^2$
term at very low temperatures (below T$_0$) from a fit
$\rho(T)=\rho_0+AT^2$ (line is a guide to the
eye).\label{QCPresistivity}}
\end{figure}
\begin{figure}
\includegraphics[width=1\hsize]{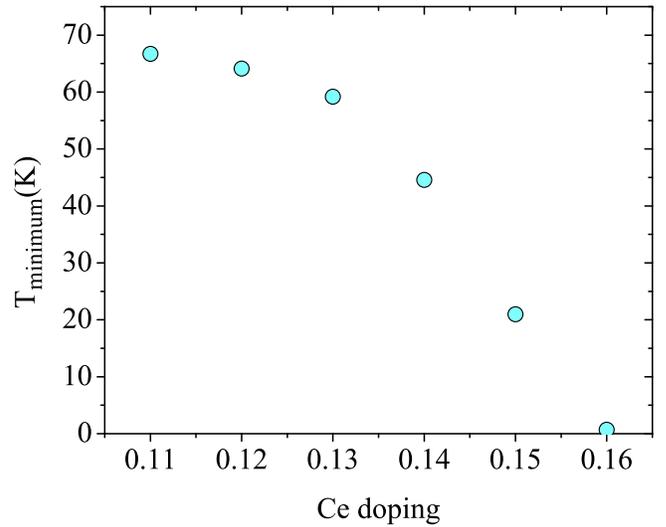}
\caption {The temperature at which minimum resistivity occurs as a
function of doping.\label{Upturn}}
\end{figure}
\par
\begin{figure}
\includegraphics[width=1\hsize]{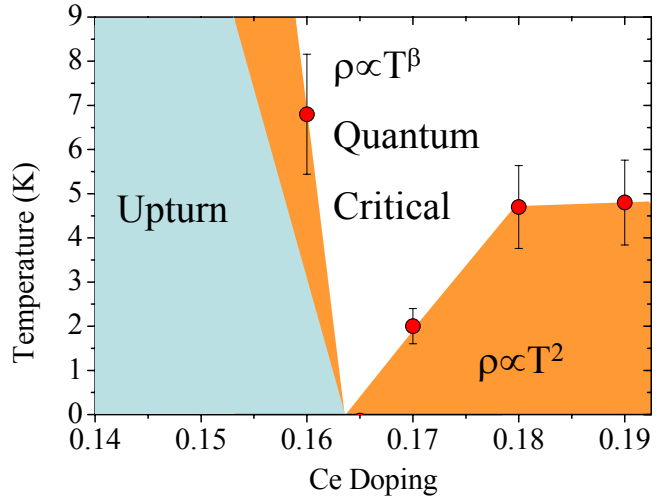}
\caption {Schematic illustration of the phase diagram of \PCCO~
from resistivity measurements. The data points are T$_0$, the
temperature below which the T$^2$ behavior is
observed.\label{phasediagram}}
\end{figure}
At very low temperatures the temperature dependence is Fermi
Liquid like, $\rho \propto $T$^2$. In Figure \ref{FLbehavioratLT}
we plot the resistivity \emph{versus} T$^2$ for the $x=0.17$
sample. T$_0$ is defined as the temperature at which the data
deviate from the linear fit. We shall discuss the slope of the
line later on. The T$^2$ region starts below T$_0=4.8K, 4.7K, 2K,$
and $6.8K \pm 5\%$ for x=0.19, 0.18, 0.17 and 0.16 respectively
(for $x\leq 0.15$ the low temperature behavior is obscured by the
upturn). Between the two T$^2$ regions we find a different
temperature dependence with exponent $1<\beta<2$.
\par
Near a \QCP~ one expects a quantum fluctuations region in the
doping-temperature phase diagram \cite{SAchdevsbook}. This region
becomes narrower as the temperature is lowered. It has a shape of
a funnel pointing towards the \QCP~. This is clearly seen, for
example, in heavy fermion materials \cite{custersQCP}. In our case
such a funnel shape region with $\beta<2$ (and maybe $\beta=1$)
with a crossover to T$^2$ behavior at low temperatures is
expected. A major experimental complication in our case is the
chemical doping as a control parameter. Such a control parameter
cannot be easily tuned as for the case of heavy fermions where the
control parameter is pressure or magnetic field.
\par
We fit the low temperature range (0.35 K to 20 K) of the
resistivity data from Figure \ref{NSresistivity} to the form $\rho
(T)=\rho_0+CT^{\varepsilon}$, with C, $\rho_0$ and $\varepsilon$
independent of temperature. The exponent, $\varepsilon$, obtained
from the fits is presented in figure \ref{QCPresistivity}. It has
a strong doping dependence and gets closer to 1 as we decrease the
Ce doping from 0.19 to 0.17. Decreasing the Ce doping further to
x=0.16 results in an increase in $\varepsilon$ to 1.4. The
behavior of the resistivity at low temperatures is consistent with
a quantum phase transition for several reasons: first, the doping
dependence of the fitting parameter, $\varepsilon$. Since we fit
our data from 0.35K to 20K we pick up the T$^2$ region at very low
temperatures and the T$^\beta$ quantum critical region. The fit is
a weighted average over these two regions. The weight of each
region changes with the distance from the \QCP~. As one approaches
the \QCP~ $\varepsilon$ should get closer to $\beta$,
$\varepsilon=\beta$ exactly at the \QCP~. The doping dependence of
$\varepsilon$ obtained from the fit over the same temperature
range for the various doping levels (Fig.\ref{QCPresistivity}a)
suggests that in a "funnel-shape" region in the doping-temperature
phase diagram a linear, or close to linear, in T resistivity
occurs. This is the behavior expected for transport properties in
the quantum critical fluctuation region at finite temperatures
above a \QCP~
\cite{custersQCP}$^,$\cite{Varmanusinovreview}$^,$\cite{Chakravartyfunnel}.
Second, the low temperature T$^2$ region below T$_0$ becomes
larger as we move away from the \QCP~. This is expected assuming
that the T$^\varepsilon$ region is governed by quantum
fluctuations. Third, in figure \ref{QCPresistivity}, we plot the
coefficient $A$ obtained when we fit the data to the form
$\rho=\rho_0+AT^2$ for the low temperature T$^2$ region, below
T$_0$, as a function of doping. Since the resistivity is
continuous, the coefficient $A$ should diverge as one approaches a
\QCP~. We find a large increase in $A$ for x=0.17, the doping at
which $\varepsilon$ has its smallest value. Finally, we note that
the temperature of minimum resistivity where an upturn appears in
$\rho(T)$ behaves in a similar way (see figure \ref{Upturn}; it
decreases with increasing doping and vanishes on the overdoped
side.
\par
The various regions of resistivity behavior in the
doping-temperature phase diagram are illustrated in figure
\ref{phasediagram}. The data points are T$_0$ the temperature
below which a T$^2$ behavior is observed as inferred from the
deviation from the linear fit in figure \ref{FLbehavioratLT}.
\par
The exponent $\beta$ at the \QCP~ itself doesn't have to be 1. But
Fournier \emph{et al.}\cite{fournierPRL} found a linear in T
resistivity from 10K down to 40mK in one of their x=0.17 \PCCO~
films, a film that had T$_c=15\pm4$ K, somewhere between our
$x=0.16$ and $x=0.17$ samples. It is possible that Fournier
\emph{et al.} hit $x_c$ in their Ce=0.17 film. Note that $x_c$
depends on both Ce and oxygen and therefore samples made by
different groups can differ slightly in Ce concentration for the
same carrier concentration. Also, as shown by A. Rosch
\cite{Rosch}$^,$\cite{RoschPRB} disorder can affect the exponent
in the \QCP~ region. Based on the value of the residual
resistivity, the films of Fournier \emph{et al.}\cite{fournierPRL}
have different disorder than ours. Taking Fournier's data into
account $\varepsilon$ appears to approach 1 around $x_c=0.165$ in
our films.
\par
In summary, above 100 K the resistivity follows similar
temperature dependence for all doping levels.\cite{DaganholeSCinElectrondoped} For $0.15<x<0.19$ we
identify a low temperature behavior characteristic of a \QPT~ at
$x_c\cong0.165$.
\subsection{\label{Hallmeasurements.} Hall measurements.}
\par
\begin{figure}
\includegraphics[width=1\hsize]{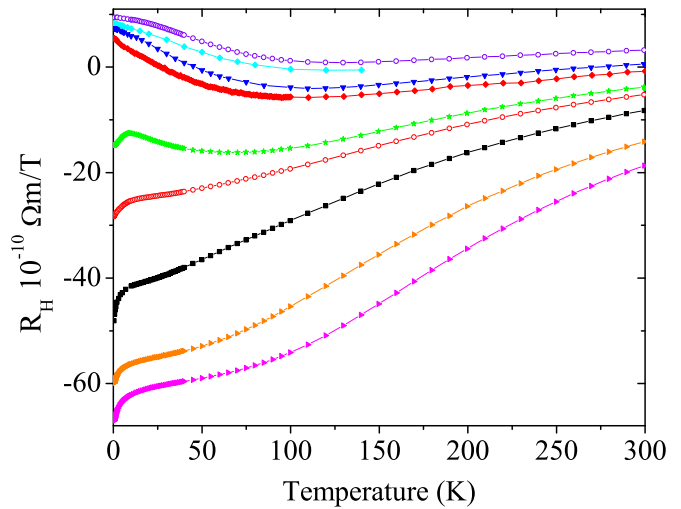}
\caption{(color online) The Hall coefficient R$_H$ as function of
temperature for the various doping levels (top to bottom):
$x=0.19$, $x=0.18$, $x=0.17$, $x=0.16$, $x=0.15$, $x=0.14$,
$x=0.13$, $x=0.12$, and $x=0.11$. \label{RHatT}}
\end{figure}

In Figure \ref{RHatT} we show
R$_H$ as a function of temperature for various doping levels. At
high temperatures (T$>$100 K) R$_H$ has a strong temperature
dependence for all doping levels. Even for the extreme ones:
x=0.19, x=0.11, R$_H$ changes by factors of 4 and 2 respectively
when the temperature is decreased from 300K to 100K. The sign of
R$_H$, which in a simple metal corresponds to the type of charge
carrier, also changes with doping. It is also changing with
temperature for $0.16\leq x \leq0.18$. The temperature dependence
of R$_H$ (along with other transport properties) was previously
interpreted as evidence for two types of carriers for Ce
concentrations near optimum doping ($x=0.15$)
\cite{fournier2bands}$^,$\cite{Golnik}. For $x<0.18$ the strong
temperature dependence of R$_H$ persists even at temperatures as
low as T$<10$ K. Armitage \emph{et al.}\cite{armitageprldoping}
have shown that the Fermi surface is changing from electron
pockets at low doping levels to a partially gapped hole like Fermi
surface at optimum doping. It is possible that the Fermi surface
completely reconnects at $x>0.17$.\cite{matsuiAPS} Indeed, we
observe no sign changes with temperature for $x=0.19$ and for
$x\leq 0.15$. However, the strong temperature dependence persists
even for our most overdoped sample ($x=0.19$) and the most
undersoped one ($x=0.11$) where the Fermi surface is supposed to
be completely hole like or electron like respectively.
\par

\begin{figure}
\includegraphics[width=1\hsize]{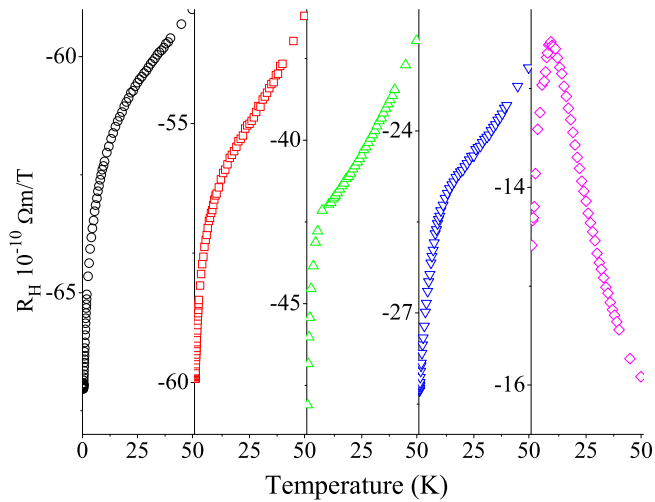}
\caption{The Hall coefficient as a function of temperatures for
various doping levels $x=0.11$(left panel), $x=0.12$, $x=0.13$,
$x=0.14$, $x=0.15$.}\label{RhatlowT}
\end{figure}

Another interesting feature seen in figure \ref{RhatlowT} for
samples with $x\leq 0.15$ is a sharp decrease in R$_H$ by about
10$\%$ below $\sim 10$ K. All these samples exhibit an upturn, an
increase in resistivity with decreasing temperature. It has been
suggested that weak localization is responsible for the upturn in
resistivity at low temperatures \cite{fournier2dWL}. However, no
first order corrections are expected for R$_H$ in the case of weak
localization \cite{leeRMP2DWL}. This suggests a decrease in the
number of carriers at low temperatures. One has to assume that
partial gapping of the Fermi surface occurs at low temperatures
for underdoped \PCCO~. The Hall number decreases by about 15$\%$
for $x=0.12-0.15$ and by about 8$\%$ for $x=0.11$ when the
temperature is decreased from 10 K to 0.35 K.

\par

\begin{figure}
\includegraphics[width=1\hsize]{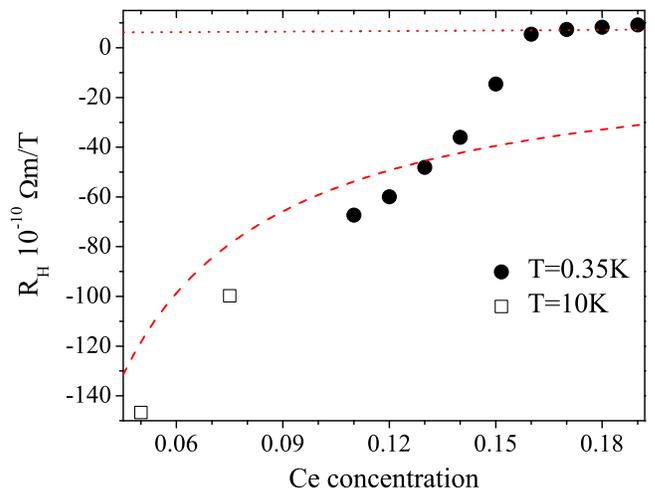}
\caption{Top: The Hall coefficient at low temperatures as a function of
doping $(x)$. Circles: T=0.35 K taken from figure \ref{RHatT},
hollow squares:T=10K data taken from ref. \cite{Onose} Dotted line
is $R_H=\frac{v_{cell}}{e(1-x)}$ and dashed line is
$R_H=-\frac{v_{cell}}{ex}$.}\label{RHQCP}
\end{figure}

In Figure \ref{RHQCP} we show R$_H$ at 0.35 K as a function of
cerium doping for $x=0.11$ to $x=0.19$ (circles). (Most data points but the x=0.14 were taken from \cite{daganResistivityPRL}The squares are
data taken from Ref.\cite{Onose}. N. P. Ong \cite{OngHall}
demonstrated that R$_H$ reflects the area swept by the mean free
path vector while going around the Fermi surface. Therefore, the
abrupt change in R$_H$ at 0.35 K is an indication of a significant
reorganization of the \FS~, which we believe may result from a
\QPT~ between two phases in the normal state
\cite{daganResistivityPRL}. From the R$_H$ behavior at 0.35 K we
identify one phase at low $x$ where R$_H$ changes rapidly and
another phase at high $x$ where R$_H$ varies more slowly. The
\QCP~ occurs where the slope of R$_H$ versus doping changes
suddenly around $x=0.165\pm 0.005$.
\par
The \QCP~ that we precisely determine from R$_H$ (and from
resistivity) is consistent with the doping trends seen in the
magnetic
\cite{LukeAFM}$^,$\cite{KangAFM}$^,$\cite{fujita:147003}$^,$\cite{Skanthakumar}
optical \cite{zimmersEPL}$^,$\cite{zimmers2007infrared} ,and ARPES
\cite{armitageprldoping} measurements. In view of the ARPES
results, the change of sign is probably due to an increase of hole
like contribution on the Fermi surface with increasing doping. But
the \QCP~ occurs presumably, when the partial gapping of the Fermi
surface disappears, which may occur at slightly higher doping. Our
$x_c$ (found from the Hall data) agrees within error with the
\QCP~ we find from the $\rho(x)$ analysis. Recently, P. Li \etal~
have found similar results from their analysis of the low
temperature thermopower in \PCCO~ \cite{PLithermopower}. We also
note that no special features are seen in either the resistivity
or the Hall coefficient at $x=0.13$, in contrast with the expected
behavior in the case of a \QCP~ at that doping as suggested by
Motoyama \etal~ \cite{GrevenQCP013}.
\par
\begin{figure}
\includegraphics[width=1\hsize]{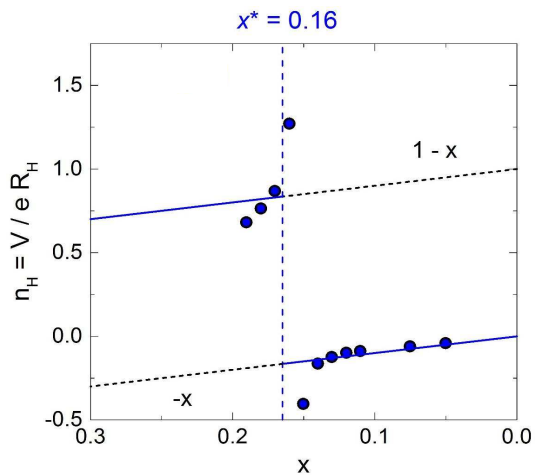}
\caption{The Hall number versus the cerium concentration per unit cell $x$ naively calculated using data from Fig.\ref{RHQCP}.}\label{nH}
\end{figure}
\par
If we naively calculate the hall coefficient,$R_H=\frac{1}{ne}$,
we do not get a simple relation to the Ce doping. For comparison,
in Figure \ref{RHQCP} the red dashed line is
$R_H=-\frac{v_{cell}}{ex}$ and the dotted line is
$R_H=\frac{v_{cell}}{e(1-x)}$ with $e$ the elementary charge
$v_{cell}$ is the unit cell volume. Onose \etal~ \cite{Onose}
pointed out that for the underdoped regime the Hall number $n_H$
behaves roughly as $x$. However, for $x=0.11$ we obtained
$n_H=0.09\pm 0.01$, different than the expected 0.11 value. On the
overdoped side $n_H$ decreases with increasing doping. The
positive sign of R$_H$ suggests a hole like Fermi surface, which
is consistent with the increasing value of R$_H$ with increasing
Ce doping and with the ARPES experiments. However, the value of
R$_H$ does not follow $\frac{1}{1-x}$ behavior as one would
naively expect from a simple electron counting even for x=0.19
where the measured Hall number is $0.64\pm 0.06$ electrons per
copper, very different from the expected value of 0.81. Lin and
Millis\cite{linandmillis} used a mean field theory to calculate
the Hall coefficient of an n-doped two dimensional system
undergoing a spin density wave transition at $x_c$.  They showed
that below $x_c$ the Fermi surface rearranges leading to changes
in R$_H$. Qualitatively our data agrees with Lin and Millis's
calculation. From our Hall data and the data of Ref.\cite{Onose}
it appears that $x_1$, the doping at which the hole pockets
disappear, is around $x_1=0.12$. However, as Lin and Millis
pointed out, the numbers obtained are rather different than the
theoretical prediction, which is not far off from the simple
$\frac{1}{x}$ and $\frac{1}{(1-x)}$ dependence as one moves away
from the critical point towards the underdoped or overdoped
regions respectively. To explain the difference between theory and
experiment they raised the possibility that the doping might be
off. In that case the $x=0.12$ sample and $x=0.19$ samples should
have "theoretical doping" of 0.01 and 0.35 respectively. This
seems unlikely since these two samples still superconduct and one
does not expect the superconducting dome to extend to such extreme
doping levels.
\par
To conclude this part, we have shown that the Hall coefficient
undergoes a change in its doping dependence at $x_c\simeq 0.165$.
This change is related to a reorganization of the Fermi surface
due to a quantum phase transition. The qualitative behavior
resembles the calculations of Lin and Millis for a density wave
scenario. Yet, a few observations remain a puzzle: the sharp
decrease of R$_H$ for $x\leq0.15$ at T$<$10 K down to 0.35 K and
the discrepancy between the measured R$_H$ at low temperatures and
the theoretical prediction away from the quantum critical region:
for $x=0.11$ and for $x=0.19$.

\subsection{\label{MR} Magnetoresistance.}

In this section we shall focus on the low temperature behavior of
the \MR~. The normal state of the n-doped cuprates is
characterized by negative \MR~ at low temperatures. Fournier
\etal~ \cite{fournier2dWL} interpreted the upturn in resistivity,
as well as the \nMR~, as a result of two dimensional weak
localization by disorder. In contrast, Sekitani \etal~
\cite{seikitani} suggested that the resistivity upturn and the
\nMR~ are due to scattering off Cu$^{2+}$ Kondo impurities induced
by residual apical oxygen. In the p-doped cuprates, for example,
negative \MR~ was found in underdoped La$_{2-x}$Sr$_x$CuO$_4$
\cite{andoboebingerfirst}. Other work was mainly focused on
in-plane \MR~ anisotropy in lightly doped La$_{2-x}$Sr$_x$CuO$_4$
\cite{Andostripes}, Pr$_{1.3}$La$_{0.7}$Ce$_x$CuO$_4$
\cite{lavrov:227003} or non superconducting
Pr$_{1.85}$Ce$_{0.15}$CuO$_{4-\delta}$ \cite{fournier:220501}. We
have recently reported \cite{daganMR} that the \nMR~ is comprised
of two contributions: spin scattering that vanishes at the \QCP~
and orbital \MR~. Here we shall elaborate on the behavior of these
two \MR~ components.
\begin{figure}
\includegraphics[width=1\hsize]{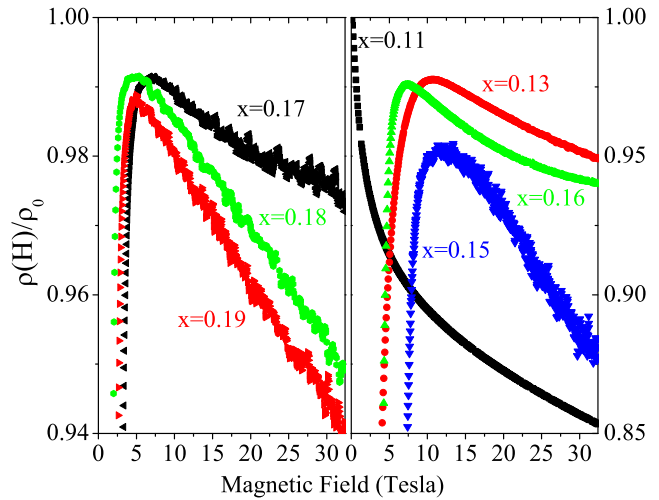}
\caption{(color online) The resistivity as a function of field at
T=1.5 K normalized with the extrapolation to zero field using a
second degree polynomial fit. The field is applied parallel to the
$c$ direction. Therefore the change in resistivity reflects both
isotropic (spin) and anisotropic (orbital) effects. \label{allMR}}
\end{figure}
\begin{figure}
\includegraphics[width=1\hsize]{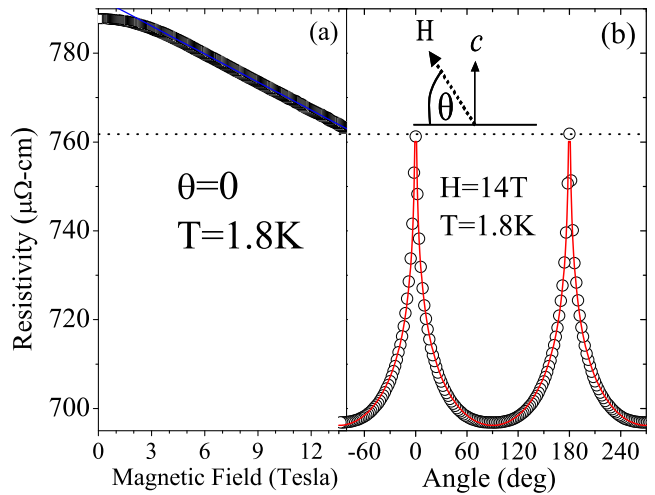}
\caption{Nonsuperconducting sample $x=0.11$ a. The resistivity as
a function of field applied parallel to the $ab$ planes. This is a
measurement of the spin \MR~. The line is a linear fit. b. The
resistivity at a constant field as a function of angle to the $ab$
plane (see insert). This is a measurement of the orbital effect.
The line is a fit to $\rho(H)=\rho(0)+b\cdot\ln|\sin(\theta)|$
\label{ORBspinMR}}
\end{figure}
\par
In figure \ref{allMR} we show the field dependence of the
normalized resistivity for various doping levels. In these
measurements the field is applied along the $c$ direction,
perpendicular to the films' surface at 1.5 K. We note a negative
magnetoresistance existing throughout the whole doping range. For
convenience the data was normalized with $\rho(0)$, the zero field
extrapolation of a polynomial fit $\rho(H)=\rho(0)+aH+bH^2$. With
$a$ and $b$ as constants. The overdoped samples (left panel)
exhibit smaller \MR~. The low field behavior is obscured by the
superconductivity. Seikitani \etal~ \cite{seikitani} have shownb
that for overdoped samples the negative \MR~ can be reduced
significantly by reducing disorder in the sample. By applying the
field along the $c$ direction we probe both the orbital and the
spin components of the \MR~. Assuming a two dimensional material
the former is sensitive only to the $c$ axis component while the
latter is approximately isotropic \cite{daganMR}. In figure
\ref{ORBspinMR} the \MR~ of the non-superconducting sample
$x=0.11$ is shown. In figure \ref{ORBspinMR}a the \MR~ is shown as
a function of field applied parallel to the $ab$ planes. Assuming
a two dimensional material, this measurement probes only the spin
magnetoresistance. The solid line is a linear fit to the data. We
note that the negative spin \MR~ is linear in field. At low
fields, $0<H<4$T it becomes quadratic in field (a linear in field
behavior at low fields would result in a singularity at $H=0$).
\par
In figure \ref{ORBspinMR}b we show the resistivity as a function
of the $c$ axis field component as the sample is rotated in a
constant field of $14$T. Since the amplitude of the field and
hence the spin \MR~ are kept constant, this measurement is
sensitive only to the orbital \MR~. We note a very strong angular
dependence of the orbital \MR~. The solid line is a fit to
$\rho(H)=\rho(0)+b\cdot\ln|\sin(\theta)|$ with $\rho(0)$ and $b$
constants and $\theta$ the angle between the film surface and the
magnetic field direction. The excellent fit suggests that the
orbital \MR~ in this sample is logarithmic with field as expected
for weak localization. However, the amplitude of the orbital \MR~
is huge (close to 10 percent) compared to the small effect
expected for a two dimensional weak localization
\cite{leeRMP2DWL}.
\par
\begin{figure}
\includegraphics[width=1\hsize]{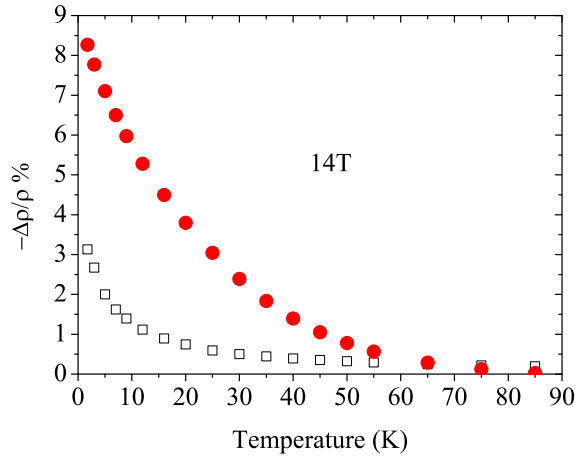}
\caption{(color online) Temperature dependence of the \MR~ for
$x=0.11$ sample at 14 T. Orbital magnetoresistance (red circles)
and spin magnetoresistance (hollow black squares).
\label{TdeporbMR}}
\end{figure}
\begin{figure}
\includegraphics[width=1\hsize]{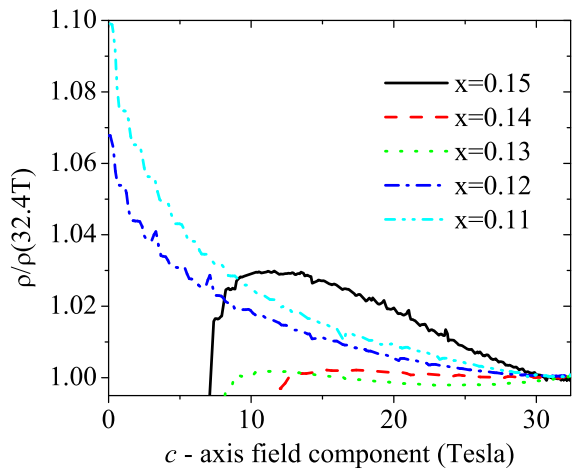}
\caption{(color online) The field dependence of the orbital \MR~
for various doping levels at 1.5K. \label{HdepOrbMR}}
\end{figure}
\begin{figure}
\includegraphics[width=1\hsize]{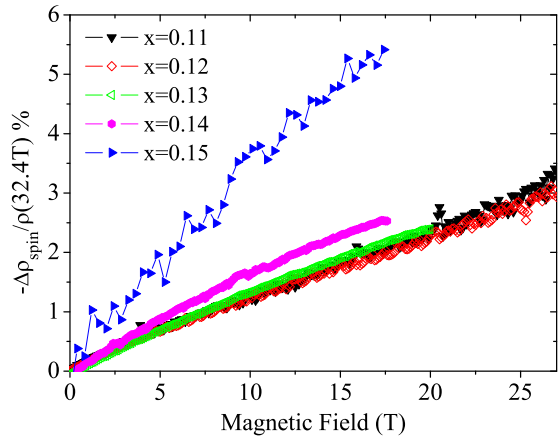}
\caption{(color online) Field dependence of the spin \MR~ for the
various doping levels at T=1.5 K. The spin \MR~ exhibits linear
field dependence even at relatively high fields. The spin \MR~
drops to zero for $x\geq0.16$ at 1.5. \label{HdepspinMR}}
\end{figure}
The temperature dependence of the orbital \MR~ $\Delta
\rho_{orb}(T)$ can be measured by subtracting out the spin \MR~
from the total one.
$\Delta\rho_{orb}(T,H)=\rho(T,H_{\|c})-\rho(T,H_{\|ab})$, with
$\rho(T,H_{\|ab})$ and $\rho(T,H_{\|c})$ the resistivity at a
certain temperature and field applied parallel to the $ab$ planes
and $c$-axis respectively. In figure \ref{TdeporbMR}
$\Delta\rho_{orb}/\rho$ at 14T is plotted as a function of
temperature (circles). The spin \MR~
$\Delta\rho_{spin}(T,H)=\rho(T,H_{\|ab})-\rho(T,H=0)$ at 14T
normalized with the zero field resistivity is also shown (hollow
squares) in figure \ref{TdeporbMR}.
\par
The non superconducting sample $x=0.11$ is the only sample where a
full rotation and parallel field dependence of the \MR~ can be
studied at low temperatures. For the superconducting samples
$x=0.12-0.19$ the effects of superconductivity persist for
parallel fields as high as 32T. For that reason the contribution
of the spin \MR~ can be inferred only from a comparison between
rotations in field and field sweeps parallel to the $c$ direction
as we did in Ref.\cite{daganMR}. In figure \ref{HdepOrbMR} the
normalized resistivity at 1.5K is plotted as a function of the $c$
axis field component. In this measurement the the sample is
rotated at a constant field of 32.4T. The total field and hence
the spin \MR~ are kept constant. This measurement is therefore
sensitive only to the orbital \MR~. For convenience the
resistivity is normalized with its value at a $c$ axis field
component of 32.4 T (the maximum value). Starting from the
underdoped sample x=0.11 the orbital \MR~ quickly decreases with
increasing doping, becoming almost zero for x=0.13, 0.14 and
increases back at x=0.15.
\begin{figure}
\includegraphics[width=1\hsize]{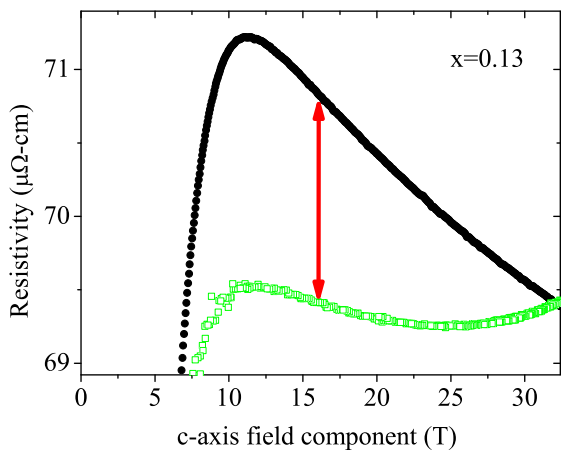}
\caption{(color online) A comparison between sweeping the field
parallel to the $c$ direction (black circles) and rotating in a
constant field of 32.4 T (green hollow squares). The difference
between these measurements is due to the spin effect (marked by an
arrow). \label{extractspinMR}}
\end{figure}
\begin{figure}
\includegraphics[width=1\hsize]{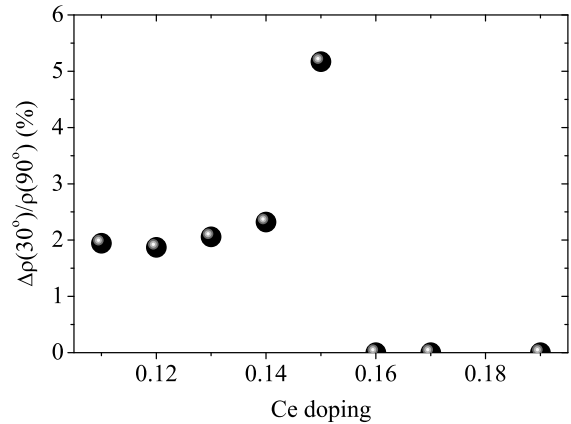}
\caption{The difference between the two measurements from figure
\ref{extractspinMR} for $c$ axis field component of 16.2 T (or
30$^\circ$). The spin \MR~ drops to zero at $x=0.16$.
\label{dopdepofSMR}}
\end{figure}
\par
To approximate the spin \MR~ for the superconducting samples we
use the following reasoning: assume that we apply a field
$\put(2,10){\vector(1,0){8}}H'=(H'_c,H'_{ab})$; the orbital \MR~
at that field can be subtracted out by measuring $\rho(H'_c,0)$.
Hence, the spin \MR~ of a field with an amplitude of
$(|\put(2,10){\vector(1,0){8}}H'|-H'_c)$ is:
$-\Delta\rho_{spin}(|\put(2,10){\vector(1,0){8}}H'|-H'_c)=\rho(H'_c,0)-\rho(\put(2,10){\vector(1,0){8}}H')$.
Here we assumed that the spin \MR~ is linear with field all the
way up to 32.4T. The spin \MR~ obtained from this procedure is
plotted for the various doping levels in figure \ref{HdepspinMR}
at 1.5 K. We note that the spin \MR~ is almost constant as a
function of doping up to $x=0.14$ then it increases at $x=0.15$
and suddenly vanishes. The most striking result found in these
measurements is the sudden disappearance of the spin \MR~ at
$x=0.16$ and 1.5 K. This result is independent of any assumption
on the field dependence of the spin \MR~. In figure
\ref{extractspinMR} we show the resistivity as a function of $c$
axis field component when the field is applied along the $c$ axis
(black circles) (in that case it is also the total applied field)
and when the sample is rotated in a constant field of 32.4 T
(hollow squares). Both measurements were taken at 1.5 K for
$x=0.13$ sample. As explained above the difference between these
two measurements is due to the spin \MR~. We arbitrarily chose a
$c$ axis field component of 16.2 T (or 30 degrees to the $ab$
planes for the case of rotation at 32.4 T) and measured the
normalized difference between these two measurements, marked in
red arrow: $\Delta\rho(30^\circ)/\rho(90^\circ)$. As we reported
previously \cite{daganMR} for doping levels $x=0.16$ and higher
the field sweep and the rotation in field show the same \MR~ as a
function of $c$ axis field component. This is because at these
doping levels the \MR~ is solely due to orbital effects. In figure
\ref{dopdepofSMR}, $\Delta\rho(30^\circ)/\rho(90^\circ)$ is
plotted as a function of doping for all the samples measured. The
spin \MR~ peaks at x=0.15 and then vanishes suddenly for overdoped
samples. The sudden disappearance of the spin \MR~ is consistent
with the existence of quantum critical point on the slightly
overdoped side.
\par
To summarize this part, a spin \MR~ exists for $x<0.16$ and the
field dependence of this \MR~ is linear at $H>2T$ and quadratic at
low fields. This rules out Kondo scattering as a reason for the
spin \MR~. In that case the \MR~ is expected to have a logarithmic
dependence on the magnetic field. It is therefore reasonable to
relate this \MR~ to the antiferromagnetic order existing at these
doping levels \cite{LukeAFM}. Scenarios such as scattering off
magnetic droplets have been proposed \cite{daganMR}, but have
received no theoretical support up to date. The sharp peak in the
spin \MR~ near optimum doping could be due to scattering off
quantum fluctuations and it is another possible evidence for a
\QPT~.
\par
The orbital \MR~ fits $\Delta\rho\propto\ln(H)$ and
$\Delta\rho\propto\ln(T)$ as predicted for weak localization.
However, this effect is rather large for $x=0.11, x=0.12$ samples;
much larger than the "standard" weak localization effect.

\section{\label{summarySec}Summary}
We have presented resistivity, Hall, and \MR~ measurements on
\PCCO~. The singular behavior of the Hall coefficient, the change
in the low temperature dependence of the resistivity and the
sudden disappearance of the spin \MR~ fit into the proposed \QCP~
scenario \cite{daganResistivityPRL}. Recent thermopower
measurements have led to similar conclusions
\cite{PLithermopower}. The nature of the phase transition cannot
be determined from our experiments, but other experiments suggest
that the transition is from an antiferromagnetic phase to a
paramagnetic one. A partial gap opening was observed by Zimmers
\etal~ in optical conductivity measurements on
\PCCO~\cite{zimmersEPL}. They interpreted their data using a spin
density wave model with a gap that vanishes at a critical doping
consistent with the \QCP~ that we report here. Muon spin rotation
measurements found that the \AFM~ phase persists up to and may be
into the superconducting dome.\cite{LukeAFM} Neutron scattering
experiments show that \AFM~ phase exists above a critical field
for optimally doped
samples.\cite{KangAFM}$^,$\cite{fujita:147003}, but no such phase
was found for overdoped samples \cite{fujita:147003}. We can
therefore assume that the transition is due to the disappearance
of a spin density wave gap followed by \FS~ rearrangement
\cite{linandmillis}. We also note that we do not find any
signatures for a \QPT~ at $x=0.13$ in our transport measurements,
in contrast with Ref.\cite{GrevenQCP013}.
\par
For the \AFM~ phase the \FS~ is electron-like at low temperatures.
It is possible that the hole-like regions seen in ARPES
\cite{armitageprldoping} do not reach the Fermi edge. The steep
decrease in R$_H$ at low temperatures ($T<20K$) for $x\leq0.15$ is
a possible evidence for this scenario. If this is indeed the case,
the difference between the Fermi edge and the hole band is of a
few meV and it is only weakly doping dependent. At higher
temperatures hole-like excitations are thermally excited and the
two types of carriers contribute to the conductivity. At the \QPT~
or slightly above it the Fermi surface becomes hole-like and its
shape changes slowly with doping (compared with the strong doping
dependence of the \FS~ in the \AFM~ phase). The agreement between
the number of carriers obtained from the low temperature Hall and
thermopower measurements by Li \etal~ \cite{PLithermopower} is
remarkable. Although Li \etal~ have used a very simple single band
model to extract the number of carriers, the obtained number is
similar to that of the Hall measurements. It is therefore possible
that even for the overdoped side ($x\geq0.16$) a single band (the
hole one) dominates at low temperatures. At higher temperatures
thermally excited electrons contribute to the complicated
temperature dependence of the Hall coefficient. However, it is
still unclear why there is such a significant difference between
the low temperature Hall coefficient and the expected theoretical
values away from the critical point\cite{linandmillis}. Other
issues that need to be resolved are: a) the origin of the peculiar
T$^2$ resistivity region at high temperatures; b) the origin of
the spin \MR~; and c) the mechanism that results in such a large
orbital \MR~.

\begin{acknowledgments}
This work was supported by NSF grant number NSF DMR-1410665 and by the Bi-National Science foundation . A portion
of this work was performed at the National High Magnetic Field
Laboratory, which is supported by NSF Cooperative Agreement No.
DMR- 0084173, by the State of Florida, and by the DOE. We thank A.
J. Millis, A. V. Chubukov, G. Blumberg, P. Li, W. Yu, J. S.
Higgins for useful discussions and M. M. Qazilbash, A. Biswas, M.
C. Barr, W. M. Fisher and R. Beck for their assistance.
\end{acknowledgments}

\bibliographystyle{apsrev}
\bibliography{Longtransport}
\end{document}